\documentclass{article}

\usepackage{microtype}
\usepackage{graphicx}
\usepackage{subfigure}
\usepackage{booktabs} 

\usepackage{hyperref}

\usepackage[accepted]{icml2025}

\usepackage{amsmath}
\usepackage{amssymb}
\usepackage{mathtools}
\usepackage{amsthm}

\usepackage{color-edits}
\addauthor{vl}{blue}

\addauthor{ZX}{orange}

\usepackage[capitalize,noabbrev]{cleveref}

\theoremstyle{plain}

\theoremstyle{definition}

\theoremstyle{remark}

\usepackage[textsize=tiny]{todonotes}

\begin{document}

\twocolumn[
\icmltitle{Rethinking Model Evaluation as Narrowing the Socio-Technical Gap}


\begin{icmlauthorlist}

\icmlauthor{Q. Vera Liao}{MSR}
\icmlauthor{Ziang Xiao}{JHU}

\end{icmlauthorlist}

\icmlaffiliation{MSR}{Microsoft Research, Montreal, Canada}
\icmlaffiliation{JHU}{Johns Hopkins University, Baltimore. USA}

\icmlcorrespondingauthor{Q.Vera Liao}{veraliao@microsoft.com}
\icmlcorrespondingauthor{Ziang Xiao}{ziang.xiao@jhu.edu}

\icmlkeywords{Machine Learning, ICML}

\vskip 0.3in
]



\printAffiliationsAndNotice{}  

\begin{abstract}
 The recent development of generative large language models (LLMs) poses new challenges for model evaluation that the research community and industry have been grappling with. While the versatile capabilities of these models ignite much excitement, they also inevitably make a leap toward homogenization: powering a wide range of applications with a single, often referred to as ``general-purpose'', model. In this position paper, we argue that model evaluation practices must take on a critical task to cope with the challenges and responsibilities brought by this homogenization: providing valid assessments for whether and how much human needs in diverse downstream use cases can be satisfied by the given model (\textit{socio-technical gap}). By drawing on lessons about improving research realism from the social sciences, human-computer interaction (HCI), and the interdisciplinary field of explainable AI (XAI), we urge the community to develop evaluation methods based on real-world contexts and human requirements, and embrace diverse evaluation methods with an acknowledgment of trade-offs between realisms and pragmatic costs to conduct the evaluation. By mapping HCI and current NLG evaluation methods, we identify opportunities for evaluation methods for LLMs to narrow the socio-technical gap and pose open questions.
\end{abstract}

\section{Introduction}
\label{intro}

The recently developed large generative models, from language models to multi-modal models, are astonishingly powerful and rapidly deployed to power diverse applications from answer engines to writing support to content analysis tools. However, these models are also giving rise to an ``evaluation crisis'', making traditional model evaluation metrics and methods inadequate or even obsolete. 

There are several hurdles to evaluating these models. First, the ML and NLP communities have long dealt with the challenges of evaluating generative models. Until recently, natural language generation (NLG) evaluation has focused on specialized models that perform tasks such as machine translation, abstractive summarization, and question-answering. These NLG models typically have open-ended and complex output spaces, and what makes the language outputs ``good'' can be multi-faceted and context-dependent. Hence, the evaluation often cannot solely rely on simple performance metrics that measure the lexical matching (e.g., ROUGE score ~\cite{lin2004rouge})with some universal ground truth. The community has developed more sophisticated metrics (e.g., BERTscore~\cite{zhang2019bertscore}) and ways to obtain ``ground truth'' (but often flawed, as criticized by a recent wave of data auditing work~\cite{blodgett2021stereotyping,raji2021ai,xiao2023evaluating}). To complement these flaws of automatic evaluation, human rated evaluation is often used in practice~\cite{zhou2022deconstructing}, such as asking people to compare or rate the quality of model outputs~\cite{sai2022survey,zheng2023judging}. However, these human evaluation practices have also been widely criticized for lacking standardization, reproducibility, and validity for assessing model utility in real-world settings~\cite{clark2021all,howcroft2020twenty,belz2021reprogen,gehrmann2022repairing}.

There is a second challenge that is more unique and critical to these ``general-purpose'' models, and one that the community only begins to grapple with: their diverse capabilities---performing all the tasks of specialized NLG model, and ``unpredictable'' behaviors---as being discovered of new ``emergent capabilities''~\cite{weiemergent}.  Initial efforts have emerged in the NLP community in the form of meta-benchmarks, by combining many specialized evaluation tasks. For example, BigBench~\cite{srivastava2022beyond} pools more than 200 tasks that were used for evaluating specialized models. Others have sought to evaluate these LLMs by human cognitive and linguistic capabilities~\cite{mahowald2024dissociating}. Amid these surging interests in model evaluation, important human-centered questions are missing: \textit{Who should care about these evaluation results?}  And \textit{how should these people make use of the evaluation results?}

In this position paper, we center our analysis on how these models will be used in practice---powering downstream applications, which must involve practitioners (e.g., product managers, developers, designers, or essentially anyone who adopts a model to create an application) to make the choice of adopting a specific model. Our analysis is also motivated by an inevitable consequence of large models that can have significant societal and ethical implications---\textit{homogenization} of using a limited number of models to power a wide range of applications. This places remarkable responsibilities on the model developers or providers to ensure their models are effective for diverse downstream use cases\footnote{We use ``use case'' to refer to a meaningful level of abstraction of how a model will be used in a cluster of applications, without limiting the term to meaning language tasks~\cite{ouyang2022training}, scenarios~\cite{liang2022holistic} or user goals~\cite{suresh2021beyond,liao2022connecting} as in prior literature. We will further discuss this as an open question in Section~\ref{open}.}, as well as clearly communicate its limitations and risks for different use cases, and enforce responsible and safe usage policies accordingly. 

In this remainder of the paper, we start by introducing the concept of \textit{socio-technical gap}, a challenge that HCI research has long contemplated regarding the inevitable divide between what a technology can do and what people need in the deployment context. We argue that the evaluation of ``general purpose'' models should especially aim to narrow this gap. We then discuss the concepts of research realism and ecological validity in the social sciences---criteria for good empirical research to ensure evidence can generalize to real-world situations. As rooted in these social science concepts, we discuss evaluation methods and considerations in HCI and the interdisciplinary area of explainable AI (XAI) to provide concrete examples and valuable lessons for closing the socio-technical gap through evaluation. Based on these lessons, \textbf{we re-frame the goal of developing model evaluation methods as narrowing the socio-technical gap along two dimensions: context realism and human requirement realism, each having possible trade-offs with pragmatic costs to conduct the evaluation}. 

While these arguments can be applied to the evaluation practices for models broadly, in Section 4, we situate our discussions within the current evaluation challenges for LLMs by mapping NLG and HCI evaluation methods along the two realism dimensions. Through this mapping, we identify new opportunities for LLM evaluation, and encourage the community to embrace diverse evaluation methods while minding their limitations and suitable situations.

\section{Socio-Technical Gap}
The concept of socio-technical gap originates from Mark Ackerman's celebrated work reflecting on the intellectual challenge faced by HCI research studying Computer-Supported Cooperative Work (CSCW)~\cite{ackerman2000intellectual}. Ackerman argues that there is an inevitable gap between the human requirements in a technology deployment context (we refer to as \textit{socio-requirements} hereafter to mean context-specific human requirements) and a given technical solution. This is because human activity is highly flexible, nuanced, and contextualized, while computational mechanisms are ``rigid and brittle'', which is inherent to the necessary formalization and abstraction. Since the socio-technical gap is inevitable, Ackerman argues that the HCI community should take it as an intellectual mission to understand this gap and ameliorate its effects. To do so, the author suggests, requires systematic exploration with ``first-order approximation''---tractable solutions that partially solve the socio-requirements while articulating their limitations and trade-offs.

\paragraph{Why is this relevant for model evaluation?} 
As computational mechanisms to be embedded in diverse social contexts, ML models will inevitably face the socio-technical gap. Especially with the current trend for homogenized use of ``general-purpose'' models, if this gap is not understood and attended to, we may produce dominant and widespread technologies that are not useful at best or even harmful to stakeholders of downstream applications. Our central proposal is that \textit{model evaluation should make a research discipline that takes up the mission of understanding and narrowing the socio-technical gap}. Developing this discipline could mean a close integration with HCI and an extension of HCI evaluation methods. We argue that this mission will require research toward the following goals: 

1) \textbf{Goal 1 (G1):} Studying people's needs, values, and activities in downstream use cases of models, and distilling principles and representations (e.g., taxonomies of prototypical use cases and socio-requirements) that can guide the evaluation methods of ML technologies. 

2) \textbf {Goal 2 (G2):} Developing evaluation methods that can provide valid and reliable assessments for whether and how much human needs in different downstream use cases can be satisfied. That is, evaluation methods should aim to be the  ``first-order approximation'' for downstream socio-requirements while articulating their limitations and trade-offs: e.g. they are \textit{proxies} for some socio-requirements, and each of them may only represent one aspect of socio-requirements. 

In the section below, we visit relevant lessons for narrowing the socio-technical gap from fields where evaluation or gathering empirical evidence has traditionally taken a central place. In particular, we focus on \textit{what} the different evaluation methods are, \textit{why} they are needed, and \textit{how} they make up proxies for socio-requirements.

\section{Lessons from Other Fields}
\label{lessons}

\subsection{Realism and Ecological Validity in the Social Sciences}

We start by lessons from what social scientists consider when designing or choosing research methods for gathering empirical evidence. Ultimately, model evaluation is about gathering evidence of how well a model works, and narrowing socio-technical gap requires the evidence to be valid for downstream use cases. In discussing strategies for ``doing research'' in social and behavioral sciences, \citet{mcgrath1995methodology} defines ``realism'' as ``the situation or context within which the evidence is gathered, in relation to the contexts to which you want your evidence to apply''. It is one of the desirable criteria to optimize for when designing research studies. For example, field studies where researchers gather evidence in natural settings such as a community or workplace is a method with the highest realism; controlled experiments tend to have lower realism by having the researchers intrude upon the situation, and a survey (even well designed) can reduce realism by obtaining participants' response out of context. McGrath also note that the criterion of realism can be (but not always) at odds with: 1) ease of having control on measurement precision, which can be considered as a form of pragmatic cost for researchers; 2) generalizability of results to other different contexts.

The concept of realism is related to \textit{ecological validity}, which more specifically focuses on how well conclusions of a laboratory study apply to the target real-world situation.~\citet{schmuckler2001ecological} outlines three dimensions of ecological validity: 1) context---how close the task or test environment is to the target real-world context; 2) human response---how well the measurement represents people's actual response and is appropriate to the constructs that matter; 3) stimuli---how well materials used in the test represent what people would encounter in the real-world context. 

All these dimensions have implications for model evaluation, even if we consider the generalizability to the real world beyond ``lab environment'' (e.g. human-subjects evaluation studies), but also human ratings and automatic metrics, which involve further abstraction of these dimensions. For example, when designing a human evaluation protocol, context realism requires the raters to be familiar with and situated in the target real-world use case when providing the ratings; and response realism requires the raters to be asked of appropriate questions reflecting the constructs that matter for the use case. When designing a benchmark, context realism can be improved by creating tasks that reflect the downstream use case; and response realism can be improved by designing metrics that reflect the constructs matter to people and how people would respond. In both cases, stimuli realism requires sampling test items that have a good coverage of what people would encounter in the given context. 


These concepts have a deep influence on how the interdisciplinary field of HCI think about designing research for evaluating how people use, interact with, and are impacted by technologies. Before turning to the lessons from the HCI field, we first focus on the area Explainable AI (XAI) at the intersection of ML and HCI, where some concrete ideas on realism of evaluation methods have been actively pursued.

\subsection{Evaluation in Explainable AI (XAI)}
Explainable AI (used here interchangeably with ``interpretable ML'') has been a surging field in the past decade, where ML, HCI and social sciences researchers have joined forces to advocate for ``human-centered evaluation''~\cite{boyd2022human,doshi2017towards}. Evaluation of XAI techniques has been taken as a unique challenge for ML for several reasons. First, different from ``traditional'' model evaluation, evaluating explainability cannot rely on ground truth, thus no existing performance metrics can be applied. Second, more importantly, the field has long recognized that explainability, by definition, is about enabling human understanding, and hence the evaluation should ideally involve human evaluation or at least reflect human desired properties of explanations. There is also a context-specific underpinning: human understanding of models is to serve some end goals. Depending on the use case, there are diverse goals people seek explanations, and hence a given XAI technique can be used for, including debugging models, taking better actions based on model outputs (e.g., improving decision outcomes when using decision-support AI systems), auditing for model biases, and more. Therefore, human-centered approaches to XAI advocate for evaluating XAI techniques according to specific downstream use cases and by their effectiveness in supporting people's end goals in the given use case~\cite{liao2021human,vaughan2020human}. 

Much of the evaluation work in XAI can be summarized under the influential ``Interpretability Evaluation Framework'' proposed by~\citet{doshi2017towards}. While recognizing ``human evaluation is essential to assessing'' XAI techniques, the authors also acknowledge that human evaluation is \textit{costly}---broadly defined, as requiring time, resources, and training that not every researcher can afford, and also costs to the human subjects such as inconvenience or even ethical concerns. To reconcile this tension, the framework proposes three levels of evaluation with increasing levels of approximation and decreasing costs:
\begin{itemize}
    \item Application-grounded evaluation: by \textit{humans} performing \textit{real tasks} with the target application. 
    \item  Human-grounded evaluation: by \textit{humans} performing \textit{simplified tasks} such as rating the ``quality'' of the explanations, which would not require the costly time, resources, and skills for completing the real task. 

    \item Functionally-grounded evaluation: without involving humans, by \textit{proxy metrics} based on some formal definitions of human desirable criteria of explanations.

\end{itemize}

This framework provides a useful viewpoint for \textbf{G2}---developing evaluation methods as valid proxies for human requirements in downstream use cases. According to the concepts of research realism and ecological validity discussed in the last section, we call out that implicit in this framework are two dimensions of proxy: proxy task for the realistic task people perform with explanation (\textit{context realism}), and proxy metric for satisfying people's requirements for the XAI techniques to be useful (\textit{human requirement realism}). We will revisit these dimensions by considering a broader set of ML and HCI evaluation methods in Section~\ref{mapping}.

The field of XAI also provides useful lessons for \textbf{G1}---studying socio-requirements in downstream use cases and distilling useful principles and representations for evaluation. To begin with, many HCI works on XAI investigated the question for context realism----\textit{what are the different use cases}, or goals and situations for people to seek model explanations~\cite{liao2020questioning,hong2020human,liao2023designerly}. For example, by synthesizing such observations from many empirical studies, ~\citet{suresh2021beyond} developed a framework that characterizes the space of people's explainability needs (i.e. different use cases of XAI) by two dimensions: people's knowledge levels and different goals. 

 This framework is informed by an important HCI lesson about ``what makes a framework useful for characterizing a technology space'', which may encompass numerous variations of applications or interfaces. ~\citet{beaudouin2004designing} proposes three criteria: \textit{descriptive power}---significant coverage of known use cases or applications; \textit{generative power}---aiding identifying new use cases or applications; and most relevant, \textit{evaluative power}---the ability to help assess solutions, including providing the structure and language to precisely define and usefully scope what is being evaluated to inform the design of evaluation practices (e.g., how to design the user study task to have context realism).

To tackle human response realism, a natural follow-up question after identifying the prototypical use cases is \textit{what are the important human requirements in these use cases}. Seeking answers to this question underlies HCI studies investigating people's needs and desires for explanations in specific applications. Recent work also began to explore this question in a more principled fashion by contrasting different use cases. For example, leveraging scenario-based survey with XAI experts and target users, \citet{liao2022connecting} systematically explored how a dozen of explanation ``goodness'' criteria (for both human evaluation and proxy metrics) appeared in XAI literature vary for different use cases of XAI, aiming to inform use-case-grounded evaluation of XAI.

\paragraph{Why is this relevant for model evaluation?} We believe these lessons for conducting human-centered and use-case-grounded evaluations are transferable for broader ML technology evaluation with the goal of narrowing the socio-technical gap. To evaluate ``large'' models that will be used to power heterogeneous applications, we encourage the community to ask: What are the prototypical use cases of these technologies? What are the human requirements in these use cases? How to develop evaluation methods that are valid proxies for these user cases and human requirements? Answering these questions will require studying people's lived experiences in specific use cases, theorizing from empirical observations to distill useful frameworks and principles, and assessing the power of these frameworks in evaluation practices, all in an iterative fashion.

\subsection{Evaluation in Human-Computer Interaction (HCI) at Large}
We now look beyond ML and AI technologies and consider the rich lessons for technology evaluation in HCI at large, which is a discipline ``concerned with the design, evaluation and implementation of interactive computing systems for human use''~\cite{hewett1992acm}. While evaluation in HCI encompasses a wide range of technologies and methodologies, we summarize a few overarching lessons by reflecting on the historical trends and perspectives in HCI evaluation. 

First, \textit{HCI embraces diverse evaluation methods for different purposes, articulating their benefits and drawbacks}. HCI has a long history of research dedicated to developing new evaluation methods. In fact, many HCI textbooks focus entirely on teaching these evaluation methods and suitable situations to use each method (e.g.,~\cite{olson2014ways}). ~\citet{barkhuus2007mice} surveyed 24 years of CHI (Conference on Human Factors in Computing Systems) literature and summarized the adopted evaluation methods along two dimensions: qualitative v.s. quantitative, and empirical (involving real users) v.s. analytical (not involving real users). They found most papers adopted empirical evaluation, with quantitative methods (e.g., lab studies, surveys) being the mainstream approaches historically, while qualitative methods (e.g., interviews, think-aloud, ethnographic studies) started catching up in the 1990s. Analytical evaluation without involving users (e.g., qualitatively done by experts, or quantitatively by user simulation models) remained to be a minority camp.

Second, \textit{the development of diverse evaluation methods is also a result of evolving views on the socio-requirements of technologies}, allowing HCI researchers to choose from methods and metrics reflecting many different aspects of socio-requirements.
~\citet{macdonald2013changing} reviewed changing perspectives in more than 70 years of evaluation practices in HCI, and summarized five phases: reliability phase (how often the system functions without failure), system performance phase (e.g., processing speed), user performance phase (user's task performance by using the system), usability phase (e.g., ease of use), and user experience phase (assessing many affective and experiential aspects beyond pragmatic functions). 

These changing perspectives are a result from the co-evolving of the types of technologies being studied---from large specialized machines operated by highly trained professionals to personal and social computing technologies that permeate many aspects of daily lives---and broadening views on how technologies can impact people instrumentally, psychologically and socially. We note a parallel with model evaluation starting from a primary focus on model reliability (e.g., accuracy) and shifting towards human-AI joint performance~\cite{lai2023towards,matias2023humans}, while currently seeing emerging socio-technical perspectives such as considering various individual, interpersonal and societal harms that can be brought about by the deployment of AI technology in complex social systems ~\cite{selbst2019fairness,shelby2023sociotechnical,weidinger2022taxonomy}. 

Lastly, \textit{HCI welcomes ``low-cost'' evaluation methods for the pragmatics, but contemplates their limitations and suitable situations}. With a pragmatic root aiming to support technology design in practice, ``costs'' to researchers and participants is another key consideration underlying the development and choices of HCI evaluation methods. For example, conducting field studies or interviews is often considered more costly of researchers' time and resources, compared to methods that gather user data more quickly and asynchronously such as survey and log analysis. 

Moreover, as discussed, HCI has historically pursued analytical evaluation methods without the necessity of conducting full-fledged user studies or even involving users at all, with a primary motivation of allowing practitioners to engage in low-cost evaluation of their designs. Acknowledging the loss of realism of nuanced interactions of real users, work on analytical evaluations aims to reduce this loss by following strong principles for approximating real users and validating such approximations. For example, there is a long line of HCI works developing cognitive models~\cite{pirolli2007cognitive,olson1995growth,card1983psychology} to simulate user interactions with one goal (among others) of providing low-cost approximation of how users would interact with an interface, and validating these models with real user interaction data. Popular analytical evaluation methods also include qualitative expert simulation (by domain experts or designers who have a deep understanding of the users) with principled protocols such as theory-based cognitive walkthrough~\cite{lewis1997cognitive} and validated heuristics for relatively universal user requirements~\cite{nielsen1995conduct}.

It is important to note that the choice of HCI evaluation methods often involves \textit{a trade-off between realism and pragmatic evaluation cost}. Under~\citet{barkhuus2007mice}'s framework, empirical evaluation by involving real participants is evidently more realistic but more costly than analytical evaluation. In many cases (but not always), qualitative methods can be more realistic but more costly (e.g. of researcher time in collecting and analyzing data)  than quantitative methods, especially with qualitative field studies. The suitable situations for choosing lower-cost, lower-realism need to be carefully justified. One commonly cited reason is ``early development phase'' where the primary goal is to filter out inferior solutions and discover issues early on. Another reason is the evaluation results being ``good enough'' for the claimed research contributions.  For example, while laboratory studies remain mainstream for HCI research making novel technical contributions that is often still at an early stage of deployment~\cite{greenberg2008usability,olsen2007evaluating}, sub-fields where the primary focus is making a positive impact for specific communities, such as  HCI4D (HCI for development), tend to embrace qualitative field studies and emphasize ``adoption, ownership and long-term use'' rather than technical usability~\cite{anokwa2009stories}.

\paragraph{Why is it relevant for model evaluation?}
We encourage model evaluation to move beyond computational metrics and embrace diverse evaluation methods in order to narrow the socio-technical gap. These evaluation methods should aim to capture diverse constructs (i.e. criteria) of socio-requirements, and allow for varying trade-offs between reflecting realistic socio-requirements and pragmatic costs to conduct the evaluation. While this view on trade-off between realism and evaluation cost resonates with ~\citet{doshi2017towards}' considerations for XAI evaluation, HCI works offer a larger set of concrete evaluation methods and valuable lessons about validating low-cost evaluation methods according to what they aim to approximate for, and articulating what are the suitable situations that justify prioritizing lowering evaluation costs.

\section{Mapping HCI and NLG Evaluation Methods for LLM Evaluation}
\label{mapping}

\begin{figure*}
    \centering
    \includegraphics[scale=0.25]{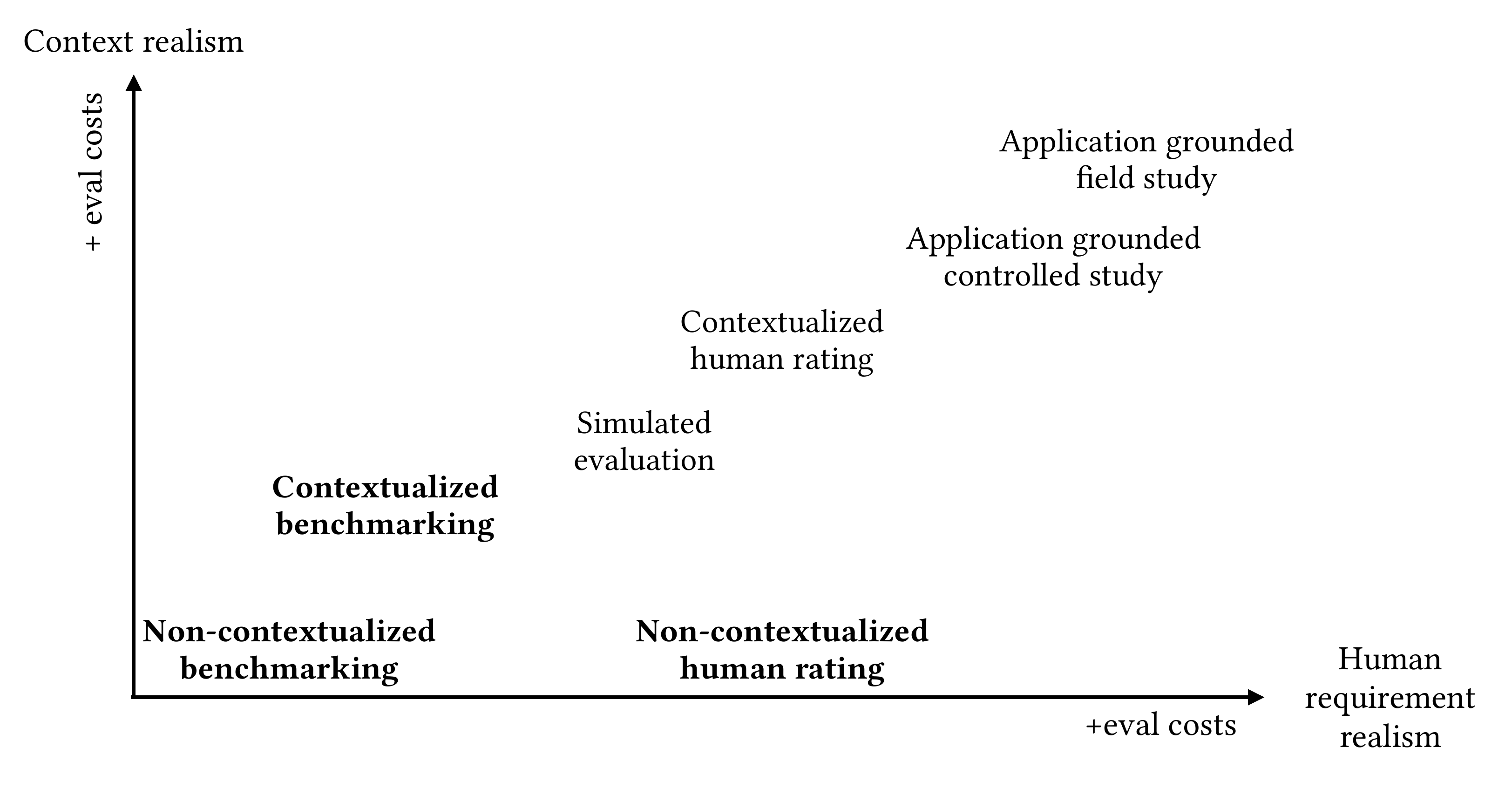}
    \caption{Mapping of HCI and NLG (in bold) evaluation methods on the two dimensions of realism}
    \label{fig:mapping}
\end{figure*}

While the lessons above can be applied to model evaluation broadly, in this section, we focus on exploring opportunities for LLM evaluations. In Figure~\ref{fig:mapping}, we map HCI evaluation methods and current NLG evaluation methods along the two dimensions of proxy for socio-requirements based on: context realism---realistic proxy for how the technology will be used in a downstream use case; and human requirement realism---realistic proxy for what requirements people involved in the use case have for the technology. We elaborate on the mapping below. This analysis is intended to re-frame different evaluation methods as different levels of proxies for narrowing the socio-technical gaps with trade-offs for pragmatic evaluation costs. With this re-framing, we highlight gaps in current LLM evaluation and identify opportunities for new methods.

\paragraph{A running example of use case} To ground our discussion, we consider a common use case of LLMs as summarizing previous social messages (emails, chats) to support future social interactions, which has already come into productization in many applications. We consider some important human requirements in this use case to be: ensuring the \textit{coverage} of important points, \textit{ease to understand} the summary, and \textit{actionability} for composing the next messages. We assume these requirements are informed by studying user needs in this use case, and that this granularity of use cases is validated to be useful for LLM evaluation (e.g., with good descriptive and evaluative power).

\textit{A few acknowledgments:} First, we note that this analysis focuses on \textbf{G2}, and is agnostic to \textit{what} the use cases are. That is, the same analysis can be performed for different downstream use cases identified, but not all evaluation methods are always applicable. Second, while we focus on the tradeoff between realism and pragmatic costs, as discussed, realism can also have tradeoffs with controllability for measurement precision, generalizability, and potentially reproducibility. Third, we take a broad, under-defined view on costs---required time, resources and training for researchers, as well as demands for human subjects, if applicable. We leave the tasks for how to further operationalize costs and make choices of evaluation methods based on costs as open questions for future work, to be discussed in the next section. Finally, this is a coarse mapping based on example methods we selectively discuss (with our own biases). It should not be taken as suggesting one category of methods is always superior to the other.

\paragraph{Non-Contextualized Benchmarking} We consider non-contextualized benchmarking as using standard NLP benchmarks for performance, including meta-benchmarks that pool these benchmarks without additional considerations for how to contextualize the interpretation of the results. These benchmarks typically measure model performance or accuracy through some definition of matching to compare the model output and the ground truth summary (reference), where the reference can be the preview sentence written by journalists to summarize the news article~\cite{DBLP:conf/nips/HermannKGEKSB15}, or a summary of 
a dialogue written by annotators with generic instruction such as ``be short...and extract important information''~\cite{gliwa2019samsum}. While typically low-cost and convenient for evaluation (e.g., many with open-source libraries available), problematically, evaluation based on these references may not reflect how people will use the model in the use case of summarizing social messages (low context realism), and accuracy to a single reference may not capture the multiple dimensions of human requirements (low human requirement realism).

\paragraph{Contextualized Benchmarking} For this category, we consider a specific example: 
a recently proposed evaluation benchmark for LLMs called HELM (Holistic Evaluation of Language Models)~\cite{liang2023holistic}. Among other contributions, HELM provides a taxonomy to map existing benchmarks by ``scenario'': e.g., the benchmark of CNN/DailyMail~\cite{DBLP:conf/nips/HermannKGEKSB15,nallapati2016abstractive} is a \textit{task} of summarization for News (\textit{domain}) in English (\textit{language}). In fact, we may consider this definition of a scenario as one way to operationalize ``downstream use cases''. This structure allows evaluation results to inform whether the LLM is good for a specific scenario, and encourages the development of new benchmarks for scenarios without appropriate benchmarks. These ``contextualized'' benchmarks, by explicitly defining the context within which an evaluation result should be interpreted, could improve context realism. Furthermore, HELM encourages ``multi-metric measurements''---in addition to accuracy, it currently evaluates all tasks by 6 other metrics such as robustness, efficiency, and biases, and supplements additional metrics that are important for specific tasks, such as factual consistency for summarization. These multi-metrics could potentially improve the human requirement realism by covering more constructs. Overall, we commend HELM as a valuable effort toward narrowing the socio-technical gap. However, we suggest that the taxonomy should be better informed by what the common downstream use cases of LLMs are, and be evaluated by its descriptive and evaluative power, rather than defined by ``tasks studied by the NLP community''. Perhaps more importantly, the choices of metrics should be based on people's actual needs and values in these use cases. Moreover, indiscriminately including all metrics for all use cases without considering their context-specific priority would not be productive for practitioners to make use of the evaluation results~\cite{liao2020questioning}.

\paragraph{Human ratings (non-contextualized v.s. contextualized)} Human ratings, by domain experts or crowd workers, are currently considered the ``gold standard'' for evaluating NLG models as they allow capturing multiple quality criteria. For example, summarization is often evaluated by fluency (grammatically correct), consistency (factual alignment with the original text), coherence (coherent body of information about a topic), etc. ~\cite{fabbri2021summeval}. Compared to automatic benchmarking, this approach can be more costly in time and resources for recruiting and training human raters.  While improving human requirement realism, we contend that this kind of normative rating, by generic constructs and rating questions (e.g., ``Are the contents of the generated text fluent?''), does not provide context realism. The results may present significant gaps in evaluating the socio-requirements in our running example. For example, the construct of \textit{actionability} may not be possible to measure without considering the context regarding when and how people will use the summary.

Our analysis reveals an opportunity to improve the context realism of human rating evaluation---developing \textit{contextualized} human rating protocols with better representations of the context of downstream use cases. For example, it could be useful to borrow from the scenario-based survey method in HCI~\cite{liao2022connecting,lubars2019ask}, which often introduces carefully designed narrative descriptions of how a system will be used to help survey takers better situate themselves to provide more realistic responses.

\paragraph{Application grounded evaluation (controlled and field studies)} Similar to~\citet{doshi2017towards}'s evaluation framework for XAI, we consider application-grounded studies to be the most costly while have the best human realism and context realism, as they allow real users to experience the system for real tasks they perform and can capture a multitude of requirements based on objective outcomes (e.g., in our running example, whether the model helps people achieve better social interactions), behavioral measures, and subjective responses (e.g., whether people report the summaries as easy to understand). This is the approach taken by the majority of HCI studies, with many guidelines and best practices to design and conduct these studies offered by HCI textbooks~\cite{olsen2007evaluating}. HCI evaluation methods further differentiate between controlled studies and field studies to evaluate technologies. Field studies are typically more costly as they often require more research resources and overhead, but they provide the best context realism and potentially better human requirement realism, allowing for uncovering and measuring constructs that may not be possible with controlled studies.

\paragraph{Simulated evaluation}Lastly, we explore a possible middle ground for balancing pragmatic costs and realism---- simulated user evaluations. While we position it in the middle of our mapping, this category may encompass a wide range of methods that may reside on different points of the realism spectrum. Simulated evaluation has also been pursued by the AI community. For example, models (whether trained by user data or top-down rules) to simulate user actions have been used to evaluate dialog models~\cite{zhang2020evaluating,griol2013automatic}, positioned as a middle ground between human- and functionally-evaluation for evaluating XAI~\cite{chenuse}, and recently spurred much interest with LLM-based user simulations~\cite{hamalainen2023evaluating,zheng2023judging,sekulic2024analysing}. By highlighting the goal of narrowing the socio-technical gap, we caution against unprincipled simulated evaluation without articulating what are being simulated and what is left out from actual user interactions \cite{agnew2024illusion}. We encourage developing \textit{use case grounded simulated evaluation}, for which useful lessons can be learned from HCI research developing cognitive models for user simulation: 1) developing the simulation with principles of user actions and target outcomes informed by empirical studies of actual users; 2) validating the simulation's approximation with real user data; 3) specifying the suitable situations to use the simulated evaluation and articulating its limitations. Lastly, informed by qualitative analytical evaluation methods in HCI~\cite{nielsen1995conduct,lewis1997cognitive}, our analysis reveals a possible opportunity for experts who have a good understanding of the target use case and users (e.g.,  UX researchers, domain experts) to evaluate the model by manually simulating the action of users, especially for more qualitative model behavioral analysis (often to reveal limitations and risks).

\section{Recommendations and Open Questions}
\label{open}

We reflect on our analysis and pose a few high-level recommendations and open questions for the community to rethink evaluation methods for LLMs and beyond.

\paragraph{Develop evaluation metrics for human-desirable constructs} While model evaluation by automatic metrics has been the dominant practice, our analysis highlights that they should be taken as a low-cost compromise to probe socio-requirements that matter in the deployment contexts. Instead of focusing on the shaky construct of ``accuracy'' to some under-defined references, the community should aim to enumerate a list of human-desirable constructs of LLM outputs by studying people's actual needs and values in downstream use cases, then develop evaluation methods accordingly, whether through scoring functions or references that are specifically designed for these constructs. A similar process should be followed to develop better-defined protocols for human ratings. Articulating and building a consensus on these constructs will also allow the community to make explicit and contemplate the values that will be encoded in the powerful technologies we develop.

\paragraph{Formalization and validation ``from the downstream up''} Our analysis reveals important connections between different types of evaluation studies: methods low on the increasing realism spectrum should be informed and validated by the methods high on the spectrum, which are closer to the downstream use case. Taking our running example, in an application-ground field study one can identify and measure the human requirements (e.g. coverage, ease to understand, actionability) for improving social interactions. Insights from the study could help inform (and use the data to validate) how to design the human rating protocol to approximate these requirements and situate the raters. Developing automatic metrics will require further formalization and abstraction.

\paragraph{What makes a useful representation of ``downstream use cases'' for LLMs?} While we adopted an under-defined concept of ``use cases'' in this paper as a ``useful level of abstraction for clusters of applications'', we believe this is an important question that requires further investigation. Besides the descriptive and generative power~\cite{beaudouin2004designing} to characterize the space of technology usage, relevant to the ``evaluative power'', one must consider the discriminative power (e.g., are the socio-requirements in different use cases sufficiently different) and the appropriate or practical level of abstraction (e.g., practical for comprehensive benchmarking across all use cases).

\paragraph{How should ``lowering LLM evaluation costs'' be defined and justified?} While we call out evaluation cost as a necessary criterion for choosing evaluation methods in practice, our analysis leaves out an important question: when and how can prioritizing a lower-cost evaluation be justified? To answer this may require unpacking types of costs, including model evaluation specific ones such as computing (including environmental impact), and further articulating ``costs'' and ``benefits'' of different evaluation methods. HCI research has considered that lower-cost evaluation can be justified by the technology development stage and claimed contributions. We believe these considerations are still applicable to model evaluation. For example, while it is acceptable for an initial academic work of a new algorithm to adopt low-cost evaluations, it would not be responsible for an organization that is making a model or a system widely available to only engage in low-cost evaluation. 

\section{Alternative Views}
A counter-argument against our position may be that model evaluation could focus on some theoretical ``intelligence'' constructs, such as language, cognitive, and reasoning capabilities. Some may draw parallels with how IQ or other standardized tests are done on people. 
However, we argue it is problematic, even irresponsible, for the field to orient around such theoretical constructs without considering how models operate in real-world contexts and, more critically, how they interact with and impact people in ways fundamentally different from human intelligence. Blindly applying those so-called ``capability constructs" could produce unuseful or even misleading evaluations.
In fact, there is a long line of education research dedicated to understanding human capabilities constructs and developing evidence-based standardized tests that are validated by how well they predict the person will succeed in future life and career. A recent paper~\cite{liu-etal-2024-ecbd} discussed applying this practice to model benchmark development by collecting evidence of success or failure from downstream use cases. Meanwhile, standardized tests have also been widely criticized for failing to capture evidence for people with diverse trajectories \cite{newman2022dropping}. We take it as a lesson for considering socio-requirements in diverse downstream use cases and developing use-case-grounded evaluation, especially for large models that will be used in heterogeneous contexts.



\bibliography{example_paper}
\bibliographystyle{icml2025}

\newpage
\appendix
\onecolumn


\end{document}